\documentclass[10pt,nofootinbib,twocolumn]{revtex4-1}

\usepackage[]{graphicx}\graphicspath{{graph/}%
}
\usepackage{amssymb,amsmath}
\usepackage{epstopdf}
\usepackage{hyperref}
\hypersetup{
  colorlinks,
  citecolor=green,
  linkcolor=blue}

\renewcommand{\Re}{{\rm Re\,}}

\renewcommand{\vec}[1]{\textnormal{\boldmath$#1$}}

\begin{document}

\title{Giant Terahertz Pulses Generated by Relativistic Beam in a Dielectric Channel}\thanks{Work supported in part by the U.S. Department of Energy under contracts No. DE-AC02-76SF00515 and DE-AC02-05CH11231.}
\author{
G. Stupakov and S. Gessner\\ SLAC National Accelerator Laboratory, Menlo Park, CA 94025, USA}
\begin{abstract}
We analyze the electromagnetic field of a short relativistic electron beam propagating in a round, hollow dielectric channel. We show that if the beam propagates with an offset relative to the axis of the channel, in a steady state, its electromagnetic field outside of the channel extends to large radii and carries an energy that scales as the Lorentz factor $\gamma$ squared (in contrast to the scaling $\ln\gamma$ without the channel). When this energy is converted into a terahertz pulse and focused on a target, the electric field in the focus can greatly  exceed typical values of the field that are currently achieved by sending beams through thin metallic foils. 

\end{abstract}
\maketitle

Terahertz radiation defined as the range between 100 GHz and 30 THz finds applications in fields as diverse as chemical and biological imaging, material science, telecommunication, semiconductor  and superconductor research \cite{Tonouchi:2007fk}. In addition to the established laser-based sources of such radiation~\cite{Auston:84,You:93}, beam-based sources utilizing short, relativistic electron bunches~\cite{Nakazato:89,Carr:02} show a great promise. To generate ultra-fast pulses of THz radiation, a short electron bunch is sent onto a thin metallic foil to emit coherent transition radiation (CTR). Experiments that use this method at the Linac Coherent Light Source (LCLS)~\cite{Daranciang:11} have obtained single-cycle pulses of radiation that is broad-band, centered on 10~THz, and contain $>0.1$~mJ of energy. In another beam-based approach, the beam passes  through a metallic pipe with a dielectric layer. As reported in~\cite{Cook:09}, using this method the authors generated narrow-band pulses with frequency 0.4~THz and energy 10~$\mu$J.

In the method that utilizes the CTR, it is the electromagnetic energy of the Coulomb field carried by the electron bunch in free space that is converted by the metallic foil to radiation. Considering as an example a spherically symmetric electron bunch with a Gaussian charge distribution moving with a relativistic  velocity $v$, it is easy to calculate its electromagnetic energy,
    \begin{align}\label{eq:1}
    W\approx(Q^2/\sqrt{\pi}\sigma)\ln\gamma
    ,
    \end{align}
where $Q$ is the bunch charge, $\sigma$ is the rms size of the bunch, and $\gamma=(1-v^2/c^2)^{-1/2}$ is the Lorentz factor. For the parameters of the experiment~\cite{Daranciang:11} where $\sigma =10\ \mu$m, $Q=0.3$ nC and $\gamma=3\times10^4$, we find $W\approx0.23$ mJ, in a reasonable agreement with the experiment. The frequency range of the CTR radiation can be estimated as $\omega\sim c/\sigma\sim 2\pi\times 2.5$ THz, which gives for the spectral energy density $dW/df\sim 2\pi W/\omega\sim 0.1$  mJ/THz.

The mechanism behind the dielectric-metal pipe is different---it's the Cherenkov radiation of the beam into waveguide modes that can propagate inside the pipe and resonantly interact with the beam. In this case, the beam generates relatively narrow-band radiation coming out from the exit of the pipe in the direction of beam propagation. The radiation energy is proportional to the length of the pipe and, in principle, can be larger than in the case of the foil. Unfortunately, the duration of the radiation pulse is long, so the energy density is low.

In both methods, the THz radiation is radially polarized, and when focussed onto a target, produces an illuminated spot with  zero field at the center, which is not optimal for most applications.

In this paper, we show how the energy of the THz pulse generated by a short, relativistic bunch can be considerably increased, replacing the scaling $W\propto\ln\gamma$ in Eq.~\eqref{eq:1} with $W\propto\gamma^2$.

\begin{figure}[htb]
\includegraphics[width=0.3\textwidth, angle=0, trim=0 0 0 0, clip]{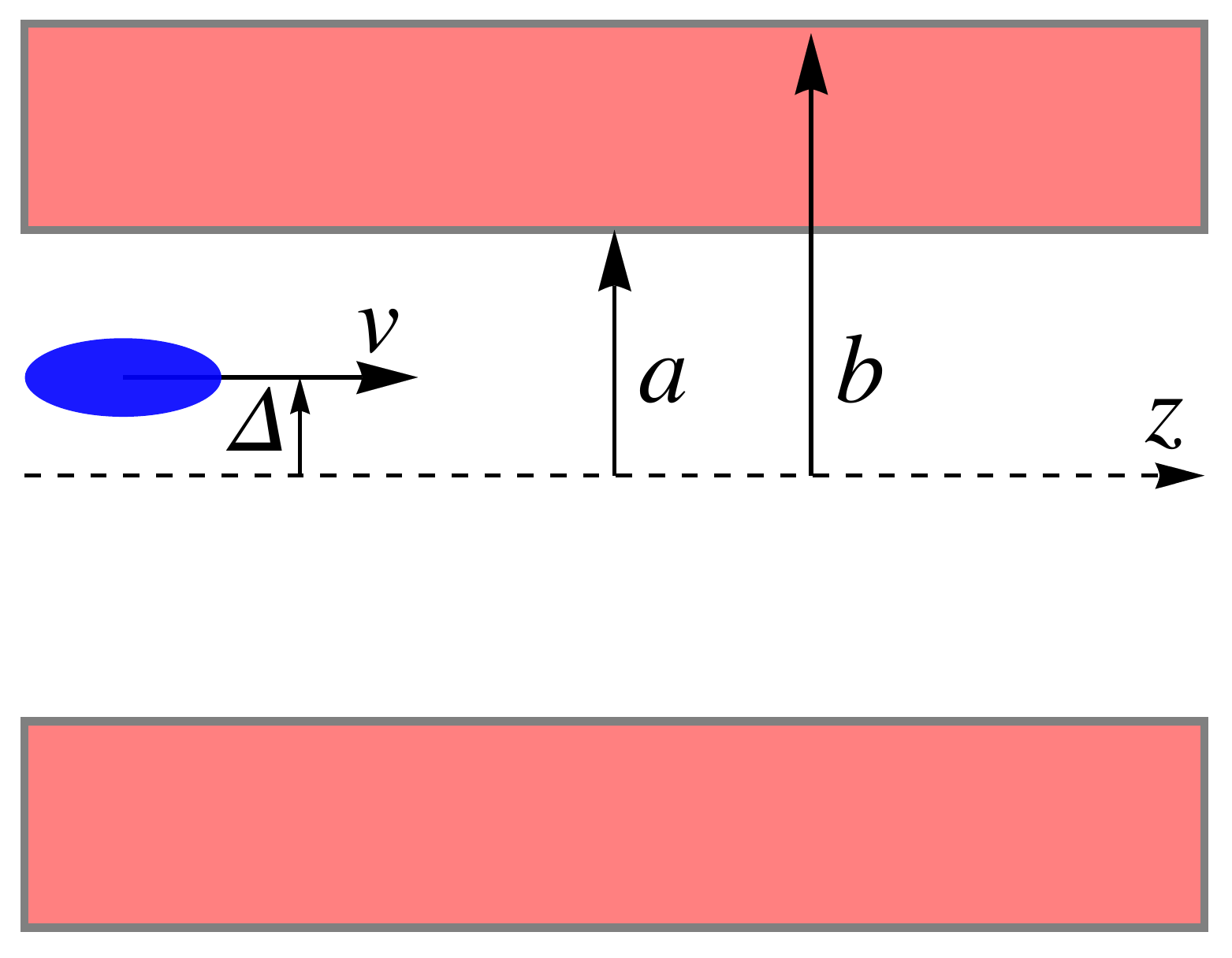}
\caption{A point charge traveling off-axis in a hollow dielectric channel shown by red color.}
\label{fig:1}
\end{figure}

We consider the setup shown in Fig.~\ref{fig:1}. 
A dielectric channel has  an inner radius $a$, an outer radius $b$, and the dielectric permittivity $\epsilon$. Outside and inside the channel $\epsilon = 1$. A short bunch of charge $Q$, which we treat as a point charge, propagates in the channel with the velocity $v$ parallel to its axis and an offset $\Delta$. Assuming $\Delta\ll a$, the electromagnetic field of the beam can be represented as a combination of the monopole axisymmetric field (corresponding to the limit $\Delta = 0$) and the dipole field proportional to $\Delta$, which is the main interest of this paper. In the cylindrical coordinate system $r,\theta,z$ centered on the axis of the channel and the angle $\theta$ measured from the direction of the offset, the dipole electric field can be represented as $\vec E(r,\theta,z,t) = \Re\tilde{\vec E}(r,z,t) e^{i\theta}$, with a similar expression for the magnetic field $\vec B$.  Our goal is to calculate these fields.

\begin{figure}[htb]
\centering
\includegraphics[width=0.45\textwidth, trim=0mm 0mm 0mm 0mm, clip]{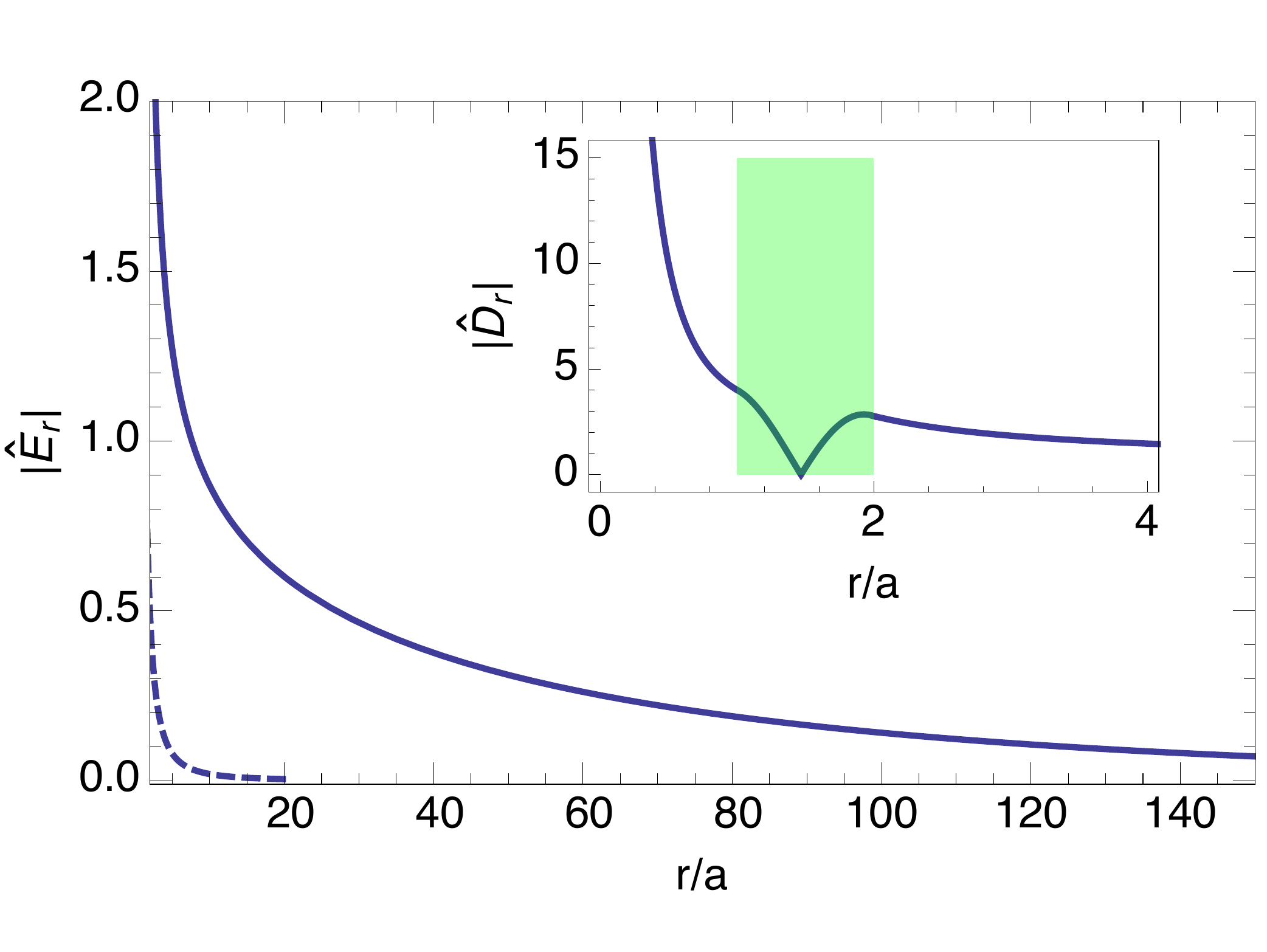}
\caption{Plot of the absolute value of the radial electric field $|\hat E_r|$ versus radius outside of the channel. The inset shows the radial distribution of $|\hat D_r|$ at small radii $r/a<4$; the region inside the dielectric is colored in green. The dashed curve near the origin shows the dipole field in free space. The fields are normalized by $p/a^2$.}
\label{fig:2}
\end{figure}

In a steady state, the fields depend on $z$ and $t$ in the combination $\xi = z-vt$, i.e., $\tilde{\vec E}(r,\xi)$. Making the Fourier transformation
    \begin{align}\label{eq:2}
    \hat{\vec E}(r,k)
    =
    \int_{-\infty}^\infty
    \tilde{\vec E}(r,\xi)
    e^{-ik\xi}
    \frac{d\xi}{2\pi}
    ,
    \end{align}
with a similar expression for the magnetic field, we substitute the fields into the Maxwell equations. The equations for $\hat{\vec E}(r,k)$ and $\hat{\vec B}(r,k)$ are then solved using the continuity of the tangential components of the fields, as well as $\epsilon \hat E_r$ and $\hat B_r$, at $r=a$ and $r=b$. These fields have a singularity at $r=0$ due to the presence of the charges and currents near the axis. The singularity is the same as in free space:
    \begin{align}\label{eq:3}
    \lim_{r\to 0}
    r^2\hat E_r
    &=
    2p
    ,\qquad
    \ \
    \lim_{r\to 0}
    r^2\hat E_\theta
    =
    -2ip
    ,\nonumber\\
    \lim_{r\to 0}
    r^2\hat B_r
    &=
    2ipv
    ,\qquad
    \lim_{r\to 0}
    r^2\hat B_\theta
    =
    2pv
    ,
    \end{align}
where $p=Q\Delta$ is the dipole moment. Using Eqs.~\eqref{eq:3} for the calculation of the dipole components of the fields, we actually treat the moving off-axis point charge as a combination of electric and magnetic moments localized on the axis of the channel, which is a valid approximation when $\Delta\ll a$. The resulting equations constitute a $4\times 4$ linear system for the boundary conditions at the dielectric-vacuum interfaces. We used the symbolic software package Mathematica to solve it analytically in the general case. An example of the absolute value of the radial electric field $|\hat E_r|$ as a function of radius obtained with this solution is shown in Fig.~\ref{fig:2} 
for the following parameters: $b/a=2$, $\epsilon = 3.8$ (dielectric constant for fused silica), $k = 2/a$ and $\gamma=200$. The field is normalized by the factor $p/a^2$. Note a remarkable feature of this field---it extends in the radial direction far beyond the channel boundaries, reaching $r\sim\gamma/k\sim 100a$. For comparison, the dashed line shows the dipole field, $|\hat E_r|=2p/r^2$ in free space (in the absence of the dielectric).

We will now focus on the structure of the field outside of the channel, at $r>b$. The electric field in this region is given by the following equations
    \begin{align}\label{eq:4}
    \hat E_r
    &=
    {iA}
    K_0
    \left(
    \rho
    \right)
    +
    \frac{iA-C\beta}{\rho}
    K_1
    \left(
    \rho
    \right)
    ,\nonumber\\
    \hat E_\theta
    &=
    {iC\beta}
    K_0
    \left(
    \rho
    \right)
    +
    \frac{A+iC\beta}{\rho}
    K_1
    \left(
    \rho
    \right)
    ,
    \end{align}
where $\rho = |k|r/\gamma$, $\beta = v/c$, and $A$ and $C$ are constant factors. The magnetic field in the limit $\gamma\gg 1$ is related to the electric one through the equations $B_r\approx -E_\theta$, $B_\theta\approx E_r$.

In the general case, the coefficients $A$ and $C$ are extremely complicated expressions involving Bessel functions. To make the problem tractable analytically, here, we consider a simplified limiting case assuming, in addition to $\gamma \gg 1$, that $b-a\ll a$ and keeping only the lowest orders in $(b-a)$ and $1/\gamma^2$. The coefficients $A$ and $C$ in this limit are given by the following formulas:
    \begin{align}\label{eq:5}
    A
    &=
    -
    \frac{2ik^2p}{\gamma^2}
    -
    iC
    ,\qquad    
    C
    =
    -
    \frac{k^2 p (\epsilon -1)^2 (b-a)}
    {a \epsilon}
    .
    \end{align}
The longitudinal components of the fields are negligibly small.

A remarkable property of this solution is that it shows that the fields outside of the channel extend in radial direction over the distance $r\sim \gamma/k$ (as we have already observed in Fig.~\ref{fig:2}) and, because of the large volume that the field occupies, it carries a considerable amount of the electromagnetic energy. To calculate the spectral energy (per unit frequency interval $d\omega=c\,dk$) we need to compute the following integral
    \begin{align}\label{eq:6}
    \frac{dW}{d\omega}
    =
    \frac{\pi}{c}
    \int_b^\infty
    rdr\,
    (|\hat E_r|^2
    +
    |\hat E_\theta|^2)
    \end{align}
where we took into account that in the relativistic limit the magnetic energy is equal to the electric one. In the limit $\gamma\to\infty$, we have $(b-a)/a\gg\gamma^{-2}$, and $A+iC\propto\gamma^{-2}$. In this case, the second terms on the right-hand side of Eqs.~\eqref{eq:4} are small and can be neglected in the calculation of the spectral energy, reducing~\eqref{eq:4} to
    \begin{align}\label{eq:7}
    \hat E_r
    \approx
    -i\hat E_\theta
    \approx
    C
    K_0
    \left(
    \rho
    \right)
    .
    \end{align}
The electric field~\eqref{eq:7} has a linear polarization along the $x$ axis (the direction of the offset) with a slow fall off in the radial direction, $\hat E_x=C    K_0    \left(    \rho    \right)$ and $\hat E_y=0$. In the limit $\gamma\to\infty$ it approaches a linearly polarized plane wave that extends to infinity.  Substituting Eq.~\eqref{eq:7} into~\eqref{eq:6} gives
    \begin{align}\label{eq:8}
    \frac{dW}{d\omega}
    =
    \frac{\pi\gamma^2}{2c k^2}
    C^2
    .
    \end{align}
We see that the spectral energy indeed scales as $dW/d\omega\propto\gamma^2$. Using this formula we can calculate the electromagnetic energy carried by a bunch of rms length $\sigma_z$: the spectrum $dW/d\omega$ should be integrated over the frequencies with the weight $e^{-\omega^2\sigma_z^2/c^2}$. As an illustrative example, consider the following set of parameters: $Q=0.3$ nC, $\sigma_z = 200\ \mu$m, $\gamma=100$, $a=0.5$ mm, $b=0.75$ mm, $\epsilon = 3.8$, $\Delta/a = 0.3$; the calculation gives $W\approx 17$ mJ. While this number is to some extent an over-estimate for the reasons that are explained below, comparing it with the estimate after Eq.~\eqref{eq:1} one sees a remarkable potential of the proposed technique.

Using Eqs.~\eqref{eq:7} we can make the inverse Fourier transform and find the electric field $E_x(r,\xi)$ of the point charge in physical space $r$, $\xi$:
    \begin{align}
    E_x(r,\xi)
    =
    -
    \pi
    \frac{p (\epsilon -1)^2 (b-a)}
    {a \epsilon}
    \frac{r^2\gamma^{-2}-2\xi^2}{(r^2\gamma^{-2}+\xi^2)^{5/2}}
    .
    \end{align}
\begin{figure}[htb]
\centering
\includegraphics[width=0.45\textwidth, trim=0mm 0mm 0mm 10mm, clip]{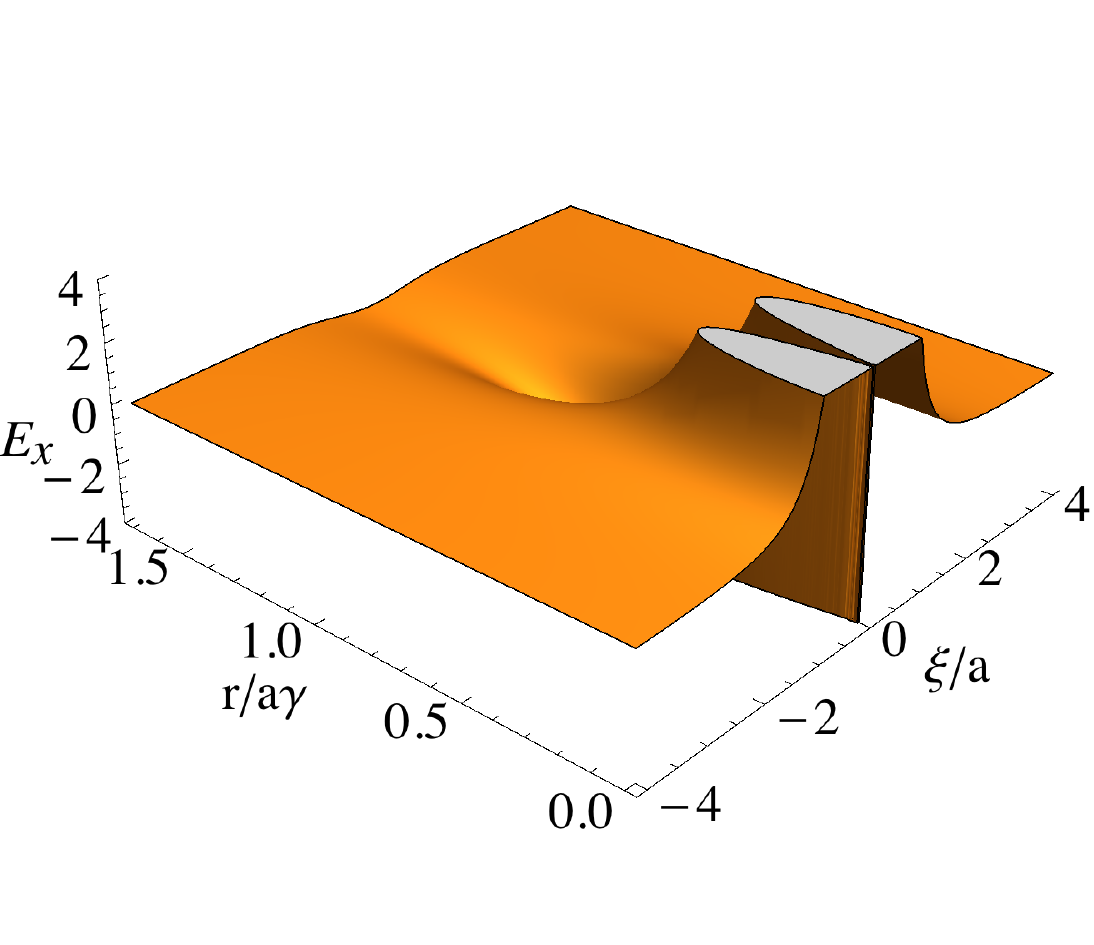}
\caption{Electric field $E_x$ (in arbitrary units) in $r-\xi$ plane around the moving dipole. Note that the radial coordinate is scaled by $\gamma a$ while the longitudinal one is scaled by $a$. The field has a singular point at the origin which is truncated in the graph. Note that the radial coordinate is scaled by $a\gamma$ and the longitudinal one by $a$.}
\label{fig:3}
\end{figure}
The plot of $E_x$ (in arbitrary units) is shown in Fig.~\ref{fig:3}; we see that the field is localized in the vicinity of the charge, $r=\xi=0$,  in a pancake-like region $\xi\sim  a$ and $ r \sim \gamma a$.

We will now explain the physical mechanism behind the giant (from $\propto\ln\gamma$ in free space to $\propto\gamma^2$ in the dielectric channel) increase in the electromagnetic energy. The total field can be considered as a superposition of the vacuum field of the moving dipole (that is the field in the absence of the dielectic) and the field generated by charges and currents in the dielectric. The vacuum field of the  dipole at small distances, $kr\ll\gamma$, is $\hat E_r\approx 2p/r^2$, $\hat E_\theta\approx-2ip/r^2$ (see Eqs.~\eqref{eq:3}). This field induces a dipole moment $dp_x(z)$ (in the direction of the beam offset) in each slice $dz$ of the dielectric tube. In the limit of a small tube thickness, $b-a\ll a$,  one can neglect the interaction between different slices taking into account only the vacuum field of the beam and integrating the polarization induced by this field, $(\epsilon-1)\vec E/4\pi$ (where $\vec E$ is the field inside the dielectric), over the volume of the slice. A straightforward calculation yields 
    \begin{align}\label{eq:9}
    dp_x=De^{ik\xi}dz
    ,
    \end{align}
where
    \begin{align}\label{eq:10}
    D
    =
    -
    \frac{p(b-a)(\epsilon-1)^2}{2\epsilon a}
    .
    \end{align}
In the next step, we calculate the radiation of these dipoles at small angles to the axes using the standard formulas of the dipole radiation with the frequency $\omega$,
    \begin{align}\label{eq:11}
    {\vec E}^\mathrm{rad}
    =
    \frac{\omega^2}{c^2R}
    e^{i\omega R/c}
    (\vec n\times d\vec p)
    \times\vec n
    ,
    \end{align}
where $\vec n = \vec R/R$, and $\vec R$ is the vector from the location of the dipole to the observation point. Recalling that $\xi=z-vt$, it follows from Eq.~\eqref{eq:9} that $\omega = kv$. With $d\vec p$ directed along $x$, the radiation field near the axis $z$ is also directed along $x$. Using $R=\sqrt{(z-z')^2+r^2}\approx |z-z'|+r^2/2|z-z'|$, where $r,z$ refer to the observation point and $z'$ is the coordinate of the slice, we arrive at the following expression for the radiation field of the dielectric tube:
    \begin{align}\label{eq:12}
    {E}^\mathrm{rad}_x
    &\approx
    \frac{\omega^2D}{c^2}
    e^{-i\omega t+ikz}
    \int_{-\infty}^z
    \frac{dz'}{|z-z'|}
    e^{{ikr^2\over 2(z-z')} - {ik(z-z')\over 2\gamma^2}}
    \nonumber\\
    &
    =
    \frac{2\omega^2D}{c^2}
    e^{-i\omega t+ikz}
    K_0(\rho)
    .
    \end{align}
With $D$ given by Eq.~\eqref{eq:10} this expression is the same as Eqs.~\eqref{eq:7} and~\eqref{eq:6}. This simple calculation corroborates our analysis based on the the direct solution of the Maxwell equations with the boundary conditions at the interfaces with the dielectric.

The mechanism described above also explains why an on-axis beam does not show the same amplification of the field energy as an off-axis one. For an on-axis beam, its electric field is axisymmetric and it does not induce dipole  moments in the slices of the dielectric. It does induce the quadrupole moments, but the quadrupole radiation is suppressed in the forward direction (along $z$), in contrast to the dipole one.

With the understanding that the field outside of the channel is generated by the radiation of the dielectric slices from the preceding part of the trajectory we can draw an important conclusion from Eq.~\eqref{eq:12} about the \emph{formation length} of the steady state field. It is easy to see that for $r\sim \gamma/k$, the integral~\eqref{eq:12} converges on the distance $z-z'\sim \gamma^2/k$. For 1 THz frequency, we have $k^{-1}\approx 50\ \mu$m and for $\gamma=100$ in our numerical example above it means that the length of the channel should be comparable, or longer, than about 0.5 m. If the channel is shorter than the formation length, while one can still achieve a considerable amplification of the electromagnetic energy in comparison with the free space, the effect will be smaller than predicted by the steady-state analysis of our paper. This is especially true when electron bunches of GeV energy ($\gamma\sim 2000$) are used in the experiment---it is impossible to reach a steady-state regime in terahertz frequency range over a reasonable length of the channel with such large value of $\gamma$. In addition, for very large values of $\gamma$, the radial extension of the field $\gamma/k$ can exceed the transverse size of the vacuum chamber in which such an experiment would be conducted, and the model of free space outside of the channel used in this work becomes invalid.

\begin{figure}[htb]
\centering
\includegraphics[width=0.35\textwidth, trim=0mm 0mm 0mm 0mm, clip]{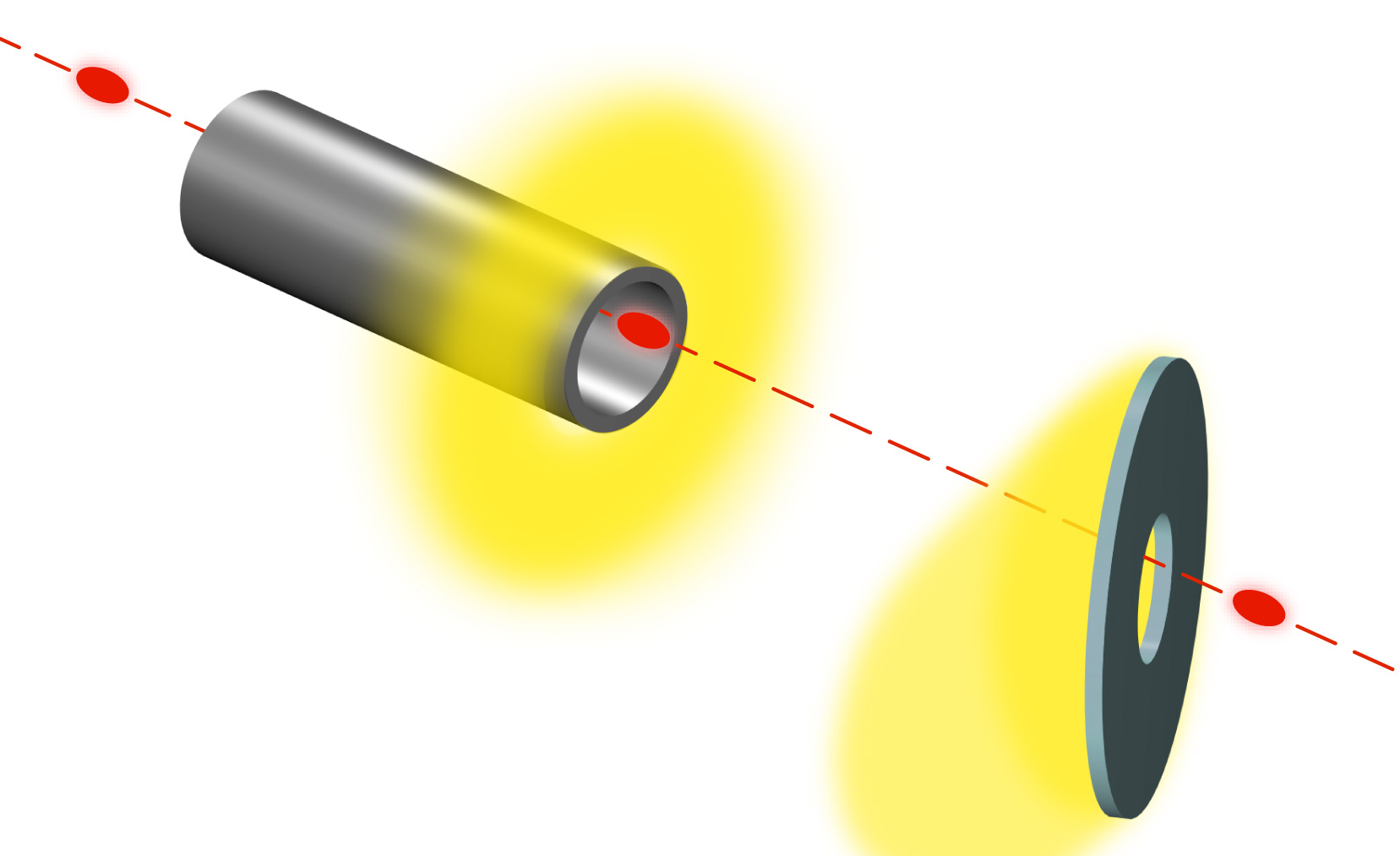}
\caption{A sketch of an experiment that generates a THz pulse from the electromagnetic field of a short relativistic bunch moving off-axis in a hollow dielectric channel. The yellow color shows the beam field extending outside of the channel. This field is intercepted by a tilted at 45 degrees metal foil and is converted into THz radiation. The radiation propagates at 90 degrees to the direction of the beam orbit.}
\label{fig:4}
\end{figure}

One more conclusion can be drawn from the previous analysis. The dipole momenta induced in the dielectric due to the offset of the beam, can be alternatively produced if the beam propagates on axis of the pipe, but the dielectric constant of the tube varies with the angle $\theta$,  $\epsilon(\theta)$. For example, the pipe can be manufactured in such a way that its top and bottom parts have different values of the dielectric constant. Such a setup may have some advantages in practice, because it facilitates the beam transport through the tube. A sketch of the experimental setup is shown in Fig.~\ref{fig:4}.

The beam propagating with an offset inside the dielectric channel experiences a strong deflecting transverse force in the direction of the offset. An estimate of this force can be easily obtained assuming $b-a\sim a$, $\epsilon-1\sim 1$ and $\Delta\sim a$,
    \begin{align}
    F_\perp
    \sim
    \frac{Q^2}{a^2}
    .
    \end{align}
For the parameters of the illustrative example after Eq.~\eqref{eq:8}, this force would deflect the beam into the wall of the dielectric at the beginning of the channel. One can try to keep the beam on the straight path applying a strong external focussing, however, a more practical approach would be to use a beam of much larger energy, on the order of several GeV. Of course, Eq.~\eqref{eq:8}, as was discussed above, is not applicable for very large $\gamma$, but an approximate energy of $dW/d\omega$ can be obtained noting that $\gamma^2/k$ in this equation is the formation length of the fields. In the limit when the the channel length $l$ is shorter than the formation length, $\gamma^2/k$ should be replaced by $l$ with the result
    \begin{align}\label{eq:8-1}
    \frac{dW}{d\omega}
    \approx
    \frac{\pi l}{2c k}
    C^2
    .
    \end{align}
Note that this formula does not involve $\gamma$ any more.

If we drop the assumption of the smallness of $b-a$ used in our analysis, numerical calculations show that, in addition to the field localized in the pancake region around the beam (as illustrated in Fig.~\ref{fig:3}), there are resonant monochromatic modes excited by the beam and propagating behind it. For example, for $b=2a$, $\epsilon = 3.8$ and $\gamma=100$, the  frequencies of the lowest four resonant modes are $\omega a/c=0.3,\ 1.3,\ 1.95,\ 2.76$. The excitation of such modes, and the electromagnetic energy that goes into them, depends on their coupling to the beam, calculations of which is beyond the scope of this paper. Conceptually, the mechanism of the excitation of these modes is the same and the experiments with the dielectric-metallic pipe of Ref.~\cite{Cook:09}, although we are dealing with the dipole modes instead of the monopole ones.

Instead of the dielectric pipe, one can use a hollow plasma channel~\cite{Gessner:2016ve}. Assuming a cold plasma, its dielectric constant is $\epsilon = 1-\omega_p^2/\omega^2$, where $\omega_p$ is the plasma frequency. In principle, the diameter of the plasma channel can be made smaller than the dielectric one, however, it may be a challenge to have a uniform, stable plasma configuration of tens of centimeters in length. In contrast to a solid dielectric, under the effect of strong relativistic fields of the bunch, the plasma response can quickly become nonlinear~\cite{Kostyukov_etal,Lu_pwfa,khudik_etal:2013}. This nonlinear plasma response, in addition to amplification of the electromagnetic energy, can, in principle, demonstrate an up-conversion of the radiation frequency when compared with the linear response of the dielectric tube.

In conclusion, in this paper we presented analysis that shows that a short relativistic beam propagating with an offset inside a dielectric channel, possesses a pancake-like electromagnetic field that extends in radial direction several orders of magnitude of the size of the channel. With the help of a standard setup using a metal foil in the path of the beam, this field can be converted into an intense, short, linearly polarized pulse of terahertz radiation. Focusing this pulse on a sample would allow to achieve record electric fields for many applications.

G.S. thanks Max Zolotorev for fruitful discussions. This work was supported by the Department of Energy, contract DE-AC03-76SF00515.


%

\end{document}